\documentclass[letter,twocolumn]{jpsj3} 
\usepackage{bm,amssymb,amsmath}	
\usepackage{graphicx}   
\usepackage{txfonts}
\newcommand{\simg}{\stackrel{>}{_\sim}}
\newcommand{\siml}{\stackrel{<}{_\sim}}
\usepackage{xcolor}

\title{
Carrier Density Dependence of Superconducting Transition Temperature in Electron-doped SrTiO$_{3}$ Based on the First-principles Calculations
}

\author{Riku Ikaida$^{1}$, Kazuhiro Sano$^{2}$, Yoshimi Masuda$^{2}$, 
Takuya Sekikawa$^{1}$, Yoshiaki \={O}no$^{1}$}

\inst{
$^1$Department of Physics, Niigata University, Niigata 950-2181, Japan \\
$^2$Department of Physics Engineering, Mie University, Mie 514-8507, Japan \\
}

\abst{
Electron-doped strontium titanate SrTiO$_{3}$, known to be one of the most dilute superconductors, is investigated on the basis of the first-principles calculations. When the carrier density $n$ decreases, the frequencies of the ferroelectric optical phonons near the $\Gamma$-point monotonically decreases in the overdoped regime with $n \simg 10^{20}/{\rm cm}^3$, while unphysical imaginary phonon frequencies due to ferroelectric instabilities appear in the underdoped regime with $n\siml 10^{20}/{\rm cm}^3$. We estimate the superconducting transition temperature $T_c$ by using the McMillan equation in the overdoped regime and find that $T_c$ increases with decreasing $n$ as consistent with experiments in the overdoped regime. Detailed analysis of the Eliashberg function reveals that the increases in $T_c$ with decreasing $n$ in the overdoped regime is mainly due to the contributions from the ferroelectric soft-mode optical phonons. 
}

\begin{document}
\maketitle
Electron-doped strontium titanate SrTiO$_3$ (STO) is known to be one of the most dilute superconductors studied extensively for more than half a century$^{1)}$. The superconductivity is induced in the insulating STO by introducing the electron carrier due to the oxygen reduction$^{2)}$ and the Nb doping$^{3)}$ for low carrier densities $n=10^{18} \sim 10^{21}/{\rm cm}^3$ corresponding to $x= 10^{-4} \sim 10^{-1}$ electrons per Ti atom. The superconducting transition temperature $T_c$ shows a characteristic dome shape as a function of $n$ with a peak of $T_c \sim 0.45$K at $n\sim 10^{20}/{\rm cm}^3$ corresponding to $x\sim 10^{-2}$\ $^{2,3)}$. In addition, a recent experiment has claimed the existence of another dome shape of $T_c$ with a peak at extremely low carrier density $n \sim 10^{18}/{\rm cm}^3$ in the case of the oxygen reduction$^{4)}$. 

In the non-doped case, the insulating STO shows a structural phase transition at $T=110$ K between the high temperature cubic phase and the low temperature tetragonal phase. The dielectric constant which is about 300 at room temperature drastically increases with decreasing $T$ below 100 K and becomes more than $2.0 \times 10^{4}$ at low temperatures, where the system is considered to be close to a ferroelectric transition$^{5)}$. In fact, the ferroelectric transition is induced by replacing $^{16}$O with $^{18}$O atoms$^{6)}$ and by replacing Sr with Ca atoms$^{7)}$. Furthermore, $T_c$ is found to be enhanced towards the quantum critical point of the ferroelectric transition$^{7, 8)}$. 

Recently, Edge {\it et al.} have proposed a phenomenological model of superconductivity in STO due to quantum ferroelectric fluctuations by taking into account of doping dependence of the ferroelectric soft-mode optical phonons shown in Fig. 1 obtained from the first-principles calculations$^{9)}$. They have found that the calculated $T_c$ well accounts for the experimental dome shape of $T_c$ for both cases with and without $^{18}$O substitution$^{9)}$. However, explicit estimates of $T_c$ based on the first-principles calculations were not performed there as unphysical imaginary phonon frequencies due to ferroelectric instabilities appear at low doping $n \siml 10^{20}/{\rm cm}^3$. 

\begin{figure}[b]
\begin{center}
\vspace{-0.1cm}
\includegraphics[width=5.0cm]{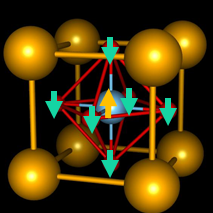}
\caption{(Color online)
Schematic figure of the ferroelectric optical phonon mode, where yellow, blue and red spheres represent Sr, Ti and O atoms, respectively. 
}
\label{Fig1}
\end{center}
\end{figure}

In this letter, we investigate the electron-doped STO on the basis of the first-principles calculations in the overdoped regime with $n=10^{20} \sim 10^{21}/{\rm cm}^3$ and obtain the carrier density dependence of the phonon dispersions including the ferroelectric soft-mode optical phonons. Based on the first-principles results with $n=2.6 \times 10^{20}/{\rm cm}^3 \sim 10^{21}/{\rm cm}^3$, where the imaginary frequencies are not observed, we estimate $T_c$ by using the McMillan equation and discuss whether the superconductivity in the overdoped STO is explained within the conventional BCS phonon mechanism. We also analyze the Eliashberg function in detail so as to clarify the contributions of the ferroelectric soft-mode optical phonons on $T_c$.

\begin{figure*}[t]
\begin{center}
\vspace{-0.2cm}
\includegraphics[width=17.0cm]{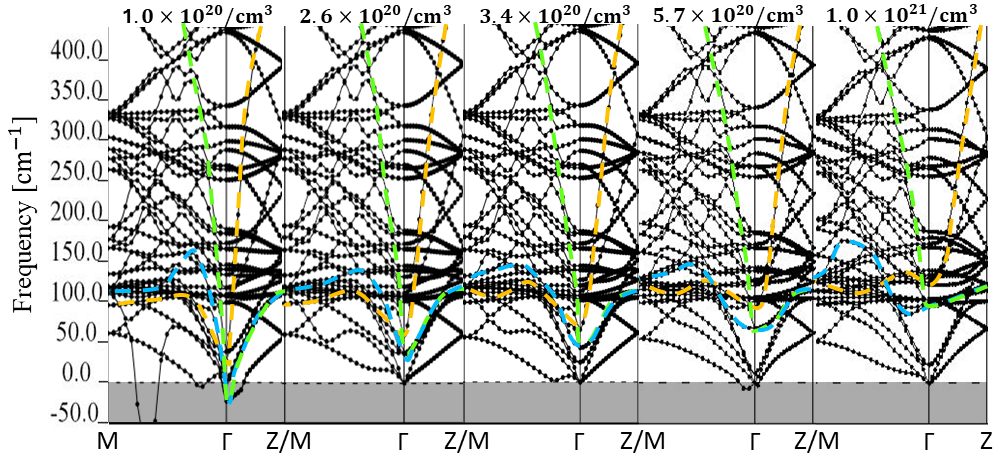}
\caption{(Color online)
Phonon dispersions on the M-$\Gamma$-Z high-symmetry line at carrier densities $n=1.0, 2.6, 3.4, 5.7 \times 10^{20}$ and $1.0 \times 10^{21}/{\rm cm}^3$, where the yellow dashed lines show branches of the ferroelectric longitudinal optical phonons and the bule and green dased lines show branches of of the ferroelectric transverse optical phonons. The imaginary phonon frequency is plotted as $-\sqrt{|\omega^2|}$ with the imaginary eigen value of $\omega^2$ in the shadow region.}
\label{Fig2}
\end{center}
\end{figure*}

In this study, the first-principles calculations are done using the Quantum ESPRESSO package, where the plane wave basis and pseudopotentials are used and the density functional theory (DFT) calculations are performed using GGA-PBE functional$^{10)}$. In the self-consistent field (SCF) calculation for electronic energy eigenvalues, we use the wave vector $\bf{k}$-mesh of 16$\times$16$\times$16. In the SCF calculation for wave functions used in phonon calculations, we use the wave vector $\bf{k}$-mesh of 8$\times$8$\times$8 for electrons and the wave vector $\bf{q}$-mesh of 4$\times$4$\times$4 for phonons so as to reduce the calculation cost$^{11)}$.  Brillouin zone integrations are performed using Monkhorst-Pack scheme$^{12)}$. The convergenece is achieved with a kenetic energy cut-off 52.0 Ry for wavefunctions and 576 Ry for charge density and potential. Gaussian type of smearing with broadening of 0.01 Ry for SCF calculations and 0.08 Ry for phonon calculations are used. 

In the electron-doped STO, the carrier density dependence is studied within the rigid-band approximation (RBA) which is a method changing only the Fermi energy of the system with doped electrons or holes to the nondoped system without transforming the band dispersions. Recent quantum oscillation experiments together with the DFT calculations using supercells have revealed that the RBA holds when the carrier is introduced by the Nb doping while it does not hold when the carrier is introduced by the oxygen reduction$^{13)}$. The deviation from the RBA in the oxygen reduction case is considered to be significant for the superconductivity at the extremely low carrier density$^{4)}$ where the superconductivity disappears in the Nb doping case. The present study, however, is restricted to the overdoped regime where $T_c$ monotonically increases with decreasing $n$ for both cases with the Nb doping and the oxygen reduction and then the RBA is expected to be applicable for describing the $n$ dependence of $T_c$ in the overdoped regime.

To obtain the stable structures in the tetragonal phase, we perform structural optimizations where the initial structures are set to be a conventional structure from the Materials Project database$^{14)}$, where the positions of O-octahedron are modifed by rotating on the $c$-axis. In the stable tetragonal phase, the 3-fold degenerate states of the botom of the conduction bands at the $\Gamma$-point in the cubic phase split into the lower 2-fold degenerate states and the upper nondegenerate state. We note that the effects of the spin-orbit coupling are not taken into account in the present study so as to reduce the heavy calculation costs to obtain the phonon dispersions and $T_c$.

Figure 2 shows the phonon dispersions below the frequency $450 \ {\rm cm}^{-1}$ on the M-$\Gamma$-Z high-symmetry line at carrier densities $n=1.0, 2.6, 3.4, 5.7 \times 10^{20}$ and $1.0 \times 10^{21}/{\rm cm}^3$. We also show branches of the ferroelectric longitudinal optical phonons (yellow dashed lines) and branches of the ferroelectric transverse optical phonons (bule and green dased lines), where the phonon mode of each branch is determined on the basis of the highest weight of the mode and several branches are connected when the branches cross each other. When $n$ decreases from $10^{21}/{\rm cm}^3$ to $2.6 \times 10^{20}/{\rm cm}^3$, the frequencies of the ferroelectric optical phonons with the both longitudinal and transverse modes near the $\Gamma$-point monotonically decrease, and then unphysical imaginary phonon frequencies with the transverse modes are observed at $n =1.0 \times 10^{20}/{\rm cm}^3$. We also calculate the phonon dispersions for cases with further low doping and find that the imaginary phonon frequencies appear for all cases with $n \siml 1.0 \times 10^{20}/{\rm cm}^3$. The results indicate that the structural instability towards the ferroelectric transition takes place at $n \sim 1.0 \times 10^{20}/{\rm cm}^3$ within the present first-principles calclations, as consistent with the previous  first-principles results$^{9)}$. On the other hand, such ferroelectric instability has not been observed in experiments of the STO except for the cases replacing $^{16}$O with $^{18}$O atoms$^{6)}$ and replacing Sr with Ca atoms$^{7)}$ mentioned before. Therefore, the effects of the ferroelectric instability are considered to be overestimated for $n \siml 1.0 \times 10^{20}/{\rm cm}^3$ in the present first-principles calculations. 

From the phonon dispersions shown in Fig. 2, we obtain the phonon frequencies at the $\Gamma$-point for the ferroelectric logitudinal mode $\omega_{\rm LO}$ and those for the lower branch of the ferroelectric transverse modes $\omega_{\rm TO}$, respectively. We plot the $n$-dependence of $\omega_{\rm LO}$ and  $\omega_{\rm TO}$ in Fig. 3, where the experimental value of $\omega_{\rm TO}$ at $n=1.1 \times 10^{20}/{\rm cm}^3$ obtained from the neutron scattering experiment$^{17)}$ is also plotted.  Although the imaginary frequencies of $\omega_{\rm TO}$ for $n \siml 1.0 \times 10^{20}/{\rm cm}^3$ contradict the experiments mentioned above, an extrapolated value from the calculated results of $\omega_{\rm TO}$ at $n=2.6, 3.4, 5.7 \times 10^{20}$ and $1.0 \times 10^{21}/{\rm cm}^3$ seems to be roughly consistent with the experimental value of $\omega_{\rm TO}$. Therefore, we expect that the carrier density dependence obtained in the present first-principles calculations is reliable for the overdoped regime with $n=2.6, 3.4, 5.7 \times 10^{20}$ and $1.0 \times 10^{21}/{\rm cm}^3$ and then use the results to estimate $T_c$ as follows.

If the superconductivity of the electron-doped STO  is explained within the conventional BCS phonon mechanism, $T_c$ can be estimated by the following McMillan equation$^{15)}$ modified by Allan and Dynes$^{16)}$
\begin{equation}
  T_{\rm{c}}=\frac{\omega_{\rm{log}}}{1.2}{\textit{\rm{exp}}}\left(-\frac{1.04(1+\lambda)}{\lambda-{\mu^{\ast}}(1+0.62\lambda)}\right)
\end{equation}
with
\begin{equation}
    {\omega_{\rm{log}}} = {\textit{\rm{exp}}}\left(\frac{2}{\lambda}{\int_{0}^{\infty}}{\frac{d\omega}{\omega}}{\alpha^{2}F(\omega)}\rm{ln}\omega \right),
\end{equation}
\begin{equation}
   {\lambda} = 2 {\int_{0}^{\infty}d\omega{\frac{\alpha^{2}F(\omega)}{\omega}}},
\end{equation}
where  $\omega_{\rm{log}}$ is the logarithmic average phonon frequency,  $\lambda$ is the electron-phonon coupling constant and $\alpha^{2}F(\omega)$ is the Eliashberg function describing the continuous spectral function of the electron-phonon interaction. In eq. (1), the Coulomb pseudopotential ${\mu^{\ast}}$ is the phenomenological parameter of order of 0.1 and we set ${\mu^{\ast}}=0.1$ in the present study. 

\begin{figure}[t]
\begin{center}
\vspace{-0.38cm}
\includegraphics[width=6.5cm]{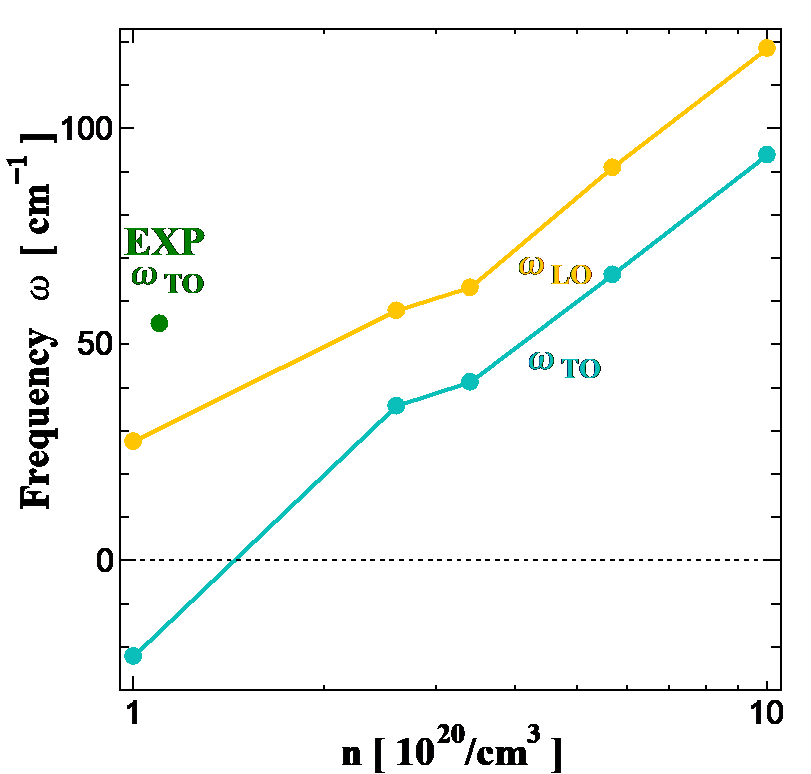}
\caption{(Color online)
The $n$-dependence of the frequencies of the ferroelectric logitudinal and transverse opical phonons at the $\Gamma$-point. The green dot shows the frequency of the ferroelectric transverse opical phonon at $n=1.1 \times 10^{20}/{\rm cm}^3$ obtained from the neutron scattering experiment$^{17)}$.  
}
\label{Fig3}
\end{center}
\end{figure}

Based on the first-principles results for the overdoped regime with $n=2.6, 3.4, 5.7 \times 10^{20}$ and $1.0 \times 10^{21}/{\rm cm}^3$, which are expected to be reliable as mentioned above, $\lambda$, $\omega_{\rm log}$ and $T_c$ are calculated using eqs. (1)-(3) and are plotted as functions of $n$ in Figs. 4 (a)-(c). When $n$ decreases, $\lambda$ monotonically increases as shown in Fig. 4 (a) that is mainly due to the enhancement of the electron-phonon interaction with the ferroelectric soft-mode optical phonons as explicitly shown later, while $\omega_{\rm log}$ monotonically decreases as shown in Fig. 4 (b) that is mainly due to the softening of the ferroelectric optical phonons as shown in Figs. 2 and 3. As the results of $\lambda$ and $\omega_{\rm log}$, $T_c$ increases with decreasing $n$ as shown in Fig. 4 (c), where the increase in $\lambda$ is more dominant as compared to the decrease in $\omega_{\rm log}$, due to the exponential (linear) dependence of $\lambda$ ($\omega_{\rm log}$) on $T_c$ in eq. (1). In Fig. 4 (c), we also plot the experimental values of $T_c$ for $\rm{SrTiO_{3-x}}$$^2)$, $\rm{Sr_{1-x}La_{x}Ti(^{16}O_{1-z} {}^{18}O_{z})_{3}}$$^8)$ and $\rm{SrTi_{1-x}Nb_{x}O_{3}}$$^4)$ and find that the $n$ dependence of the calculated values of $T_c$ seems to be consistent with that of the experimens, although the values of $T_c$ largely depend on the experiments.

\begin{figure}[t]
\begin{center}
\vspace{-0.44cm}
\includegraphics[width=5.5cm]{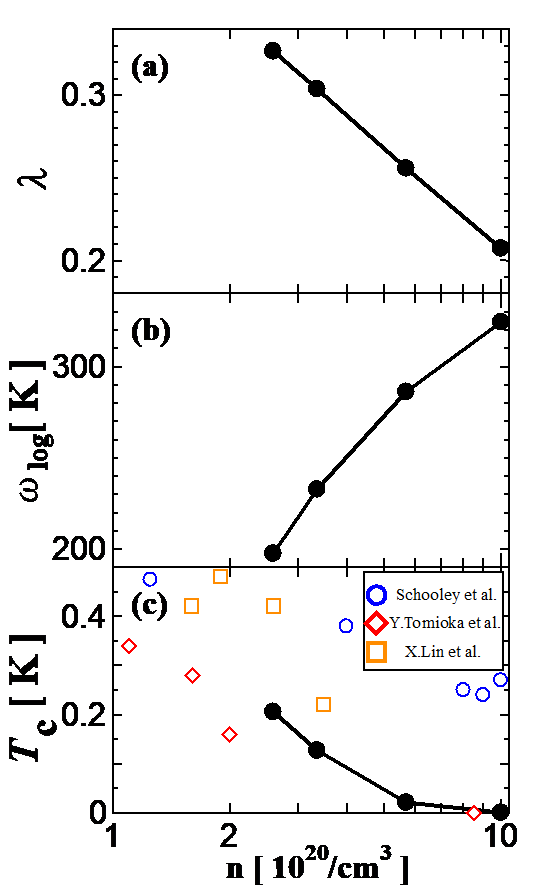}
\caption{(Color online)
The electron-phonon coupling constant $\lambda$ (a), the logarithmic average phonon frequency $\omega_{\rm log}$ (b) and the superconducting transition temperature $T_c$ (c) as functions of $n$, where the experimental values of $T_c$ for $\rm{SrTiO_{3-x}}$$^2)$ (blue circles), $\rm{Sr_{1-x}La_{x}Ti(^{16}O_{1-z} {}^{18}O_{z})_{3}}$$^8)$ (red rhombuses) and $\rm{SrTi_{1-x}Nb_{x}O_{3}}$$^4)$ (yellow squares) are also plotted in (c). 
}
\label{Fig4}
\end{center}
\end{figure}

In order to clarify the contributions of the ferroelectric soft-mode optical phonons on $\lambda$ and $T_c$, we analyze the $\omega$-dependence of $\alpha^2 F(\omega)$ together with the electron-phonon coupling integral $\lambda(\omega)$ defined by
\begin{equation}
   {\lambda(\omega)} = 2 {\int_{0}^{\omega}d\omega'{\frac{\alpha^{2}F(\omega')}{\omega'}}}.
\end{equation}
Figs. 5 (a)-(d) show the $\omega$-dependence of $\alpha^{2}F(\omega)$ and $\lambda(\omega)$ at $n=2.6, 3.4, 5.7 \times 10^{20}$ and $1.0 \times 10^{21}/{\rm cm}^3$. We see that $\alpha^2 F(\omega)$ shows significant three peaks around 100 $\rm{cm}^{-1}$, 500 $\rm{cm}^{-1}$ and 750 $\rm{cm}^{-1}$. Simultaneously, $\lambda(\omega)$ shows considerable increases around these frequencies.  Among them, the lowest peak around $100 \rm{cm}^{-1}$, which is due to the ferroelectric soft-mode optical phonons as shown in Fig. 2, mainly responsible for the increase in $\lambda$ with decreasing $n$ as shown in Fig. 5. Therefore, we conclude that the increases in $\lambda$ and $T_c$ with decreasing $n$ in the overdoped regime are mainly due to the contributions from the ferroelectric soft-mode optical phonons.

To elucidate the contribution of the electron-phonon coupling to $\lambda(\omega)$, the phonon DOS, ${\rm DOS_{ph}}(\omega)$, and the integrated phonon DOS, $\int_{0}^{\omega} \frac{\rm{DOS_{ph}}(\omega')}{\omega'}d\omega'$ are also calculated and are found to be almost independent of $n$ as shown in Fig. 6. We note that the effects of the $n$-dependence of the ferroelectric soft-mode optical phonons on ${\rm DOS_{ph}}(\omega)$ are not visible because lots of the other phonon branches are almost independent of $n$ as shown in Fig. 2. Comparing Fig. 5 with Fig. 6, we find that the increase in $\lambda(\omega)$ with decreasing $n$ is mainly due to the increase in the electron-phonon coupling for the ferroelectric soft-mode optical phonons.

\begin{figure}[t]
\begin{center}
\vspace{-0.22cm}
\includegraphics[width=6.8cm]{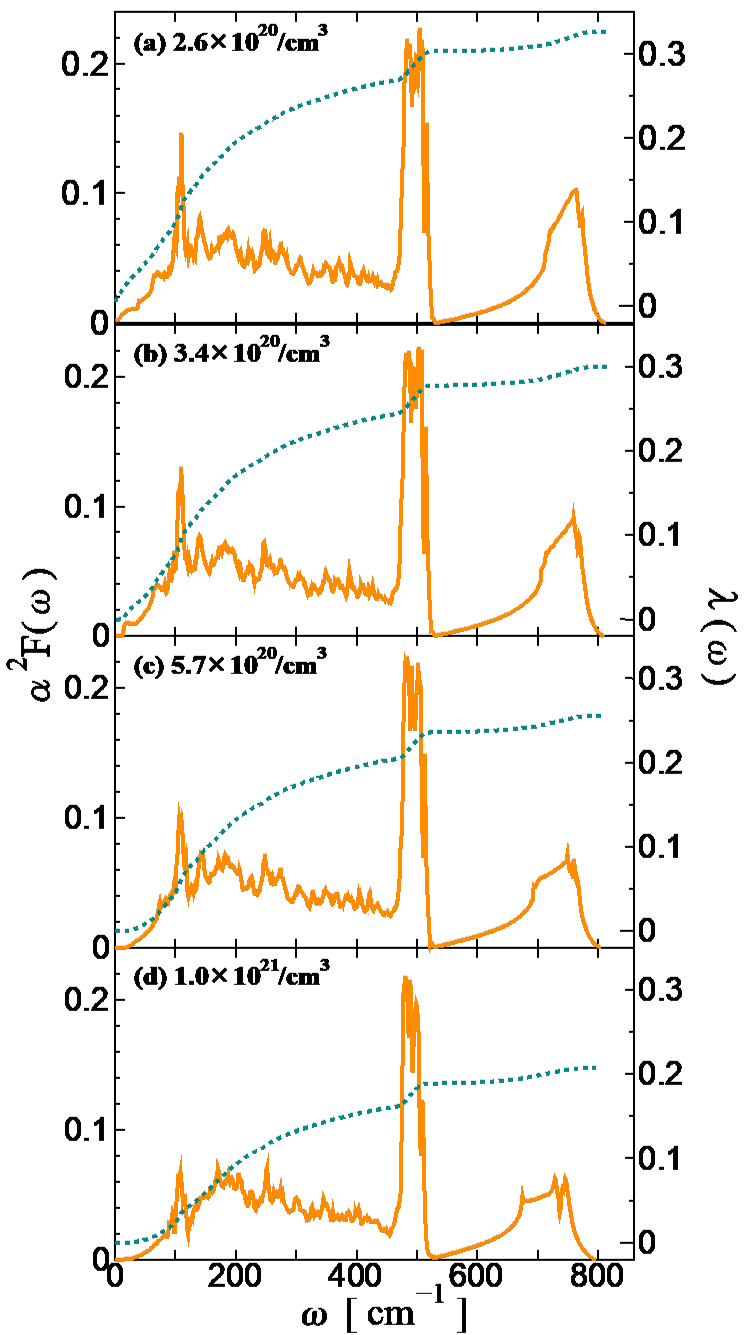}
\caption{(Color online)
The $\omega$-dependence of the Eliashberg function $\alpha^2F(\omega)$ (yellow solid lines) and the electron-phonon coupling integral $\lambda(\omega)$ (blue dotted lines) at $n=2.6 \times 10^{20}/{\rm cm}^3$ (a), $3.4 \times 10^{20}/{\rm cm}^3$ (b), $5.7 \times 10^{20}/{\rm cm}^3$ (c) and $1.0 \times 10^{21}/{\rm cm}^3$ (d). 
}
\label{Fig5}
\end{center}
\end{figure}

In summary, we have investigated the electron-doped STO on the basis of the first-principles calculations (Quantum ESPRESSO). When the carrier density $n$ decreases, the frequencies of the ferroelectric optical phonons near the $\Gamma$-point monotonically decreases in the overdoped regime with $n \simg 2.6 \times 10^{20}/{\rm cm}^3$, while unphysical imaginary phonon frequencies due to ferroelectric instabilities appear in the underdoped regime with $n\siml 1.0 \times 10^{20}/{\rm cm}^3$. We have estimated $T_c$ by using the McMillan equation in the overdoped regime and have found that $T_c$ increases with decreasing $n$ as consistent with experiments in the overdoped regime. We have also analyzed the Eliashberg function and have found that the increase in $T_c$ with decreasing $n$ is mainly due to the contributions from the ferroelectric soft-mode optical phonons. 

\begin{figure}[t]
\begin{center}
\vspace{-0.19cm}
\includegraphics[width=8.0cm]{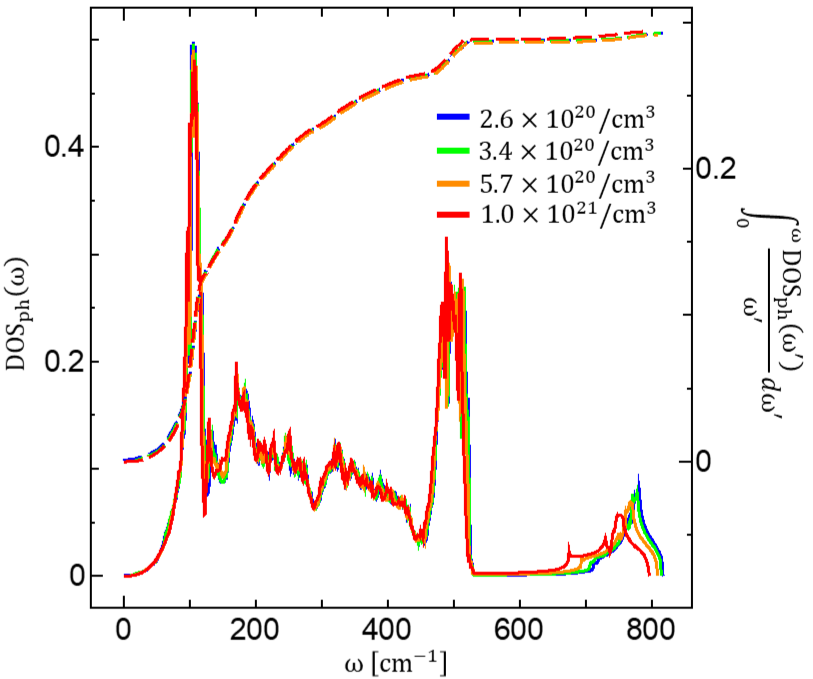}
\caption{(Color online)
The $\omega$-dependence of the phonon DOS, $\rm{DOS_{ph}}(\omega)$ (solid lines), and the integrated phonon DOS, $\int_{0}^{\omega} \frac{\rm{DOS_{ph}}(\omega')}{\omega'}d\omega'$ (dashed lines). 
}
\label{Fig6}
\end{center}
\end{figure}

In the present study, $T_c$ has not been calculated in the underdoped regime with $n\siml 1.0 \times 10^{20}/{\rm cm}^3$ where the imaginary phonon frequencies appear. In order to prevent the imaginary phonon frequencies near the $\Gamma$-point, we have also employed the tetrahedron method instead of the Gaussian broadening and have obtained some preliminary results of $T_c$ which increases with decreasing $n$ as consistent with the present results from the Gaussian broadening in the overdoped regime and continues to increase further in the lower doped region down to $n\sim 1.0 \times 10^{20}/{\rm cm}^3$. Detailed calculations with the tetrahedron method are now under way and explicit results will be reported in a subsequent paper.

\begin{acknowledgments}
This work was partially supported by JSPS KAKENHI Grant Number 25K07201 and JST SPRING, Grant Number JPMJSP2121. Numerical calculations were performed using the facilities belonging to Center for Computational Materials Science, Institute for Materials Research, Tohoku University, the Center for Computational Sciences, University of Tsukuba and Research Center for Computational Science, Okazaki, Japan. 
\\
\hrulefill 
\end{acknowledgments}

\footnotesize \ \ \ \ \ \ E-mail : r.ikaida.phys.@gmail.com

\footnotesize \ \ 1) For a review, see M. N. Gastiasoro, J. Ruhman, R. M. Fernandes, Ann. 
\\ \ \ \ \ \ \ \ \ \ \ \ Phys. \textbf{417}, 168107 (2020).

\footnotesize \ \ 2) J. F. Schooley, W. R. Hosler and M. L. Cohen, Phys. Rev. Lett. \textbf{12}, 474 
\\ \ \ \ \ \ \ \ \ \ \ \ (1964).

\footnotesize \ \ 3) C. S. Koonce, M. L. Cohen, J. F. Schooley, W. R. Hosler and E. R. Pf-
\\ \ \ \ \ \ \ \ \ \ \ \ eiffer, Phys. Rev. \textbf{163}, 380 (1967).

\footnotesize \ \ 4) X. Lin, G. Bridoux, A. Gourgout, G. Seyfarth, S.Kr\"{a}mer, M. Nardone, 
\\ \ \ \ \ \ \ \ \ \ \ \ B. Fauqu\'{e} and K. Behnia, Phys. Rev. Lett. \textbf{112}, 207002 (2014).

\footnotesize \ \ 5) E. Sawaguchi, A. Kikuchi and Y. Kodera, J. Phys. Soc. Jpn. (1962).

\footnotesize \ \ 6) M. Itoh, R. Wang, Y. Inaguma, T. Yamaguchi, Y-J. Shan and T. Naka-
\\ \ \ \ \ \ \ \ \ \ \ \ mura, Phys. Rev. Lett. \textbf{82}, 3540 (1999).

\footnotesize \ \ 7) C. Rischau, X. Lin, C. Grams, D. Finck, S. Harms, J. Engelmayer, T. 
\\ \ \ \ \ \ \ \ \ \ \ \ Lorenz, Y. Gallais, B. Fauque, J. Hemberger and K. Behnia, Nat. 
\\ \ \ \ \ \ \ \ \ \ \ \ Phys. \textbf{13}, 643 (2017).

\footnotesize \ \ 8) Y. Tomioka, N. Shirakawa, K. Shibuya and I. H. Inoue, Nature 
\\ \ \ \ \ \ \ \ \ \ \ \ Commun. \textbf{10}, 738 (2019).

\footnotesize \ \ 9) J. M. Edge, Y. Kedem, U. Aschauer, N. A. Spaldin and A. V. Balatsky, 
\\ \ \ \ \ \ \ \ \ \ \ \ Phys. Rev. Lett. \textbf{115}, 247002 (2015).

\footnotesize10) P. Giannozzi, S. Baroni, N. Bonini, M. Calandra, R. Car, C. Cavazzo-
\\ \ \ \ \ \ \ \ \ \ \ \  ni, D. Ceresoli, G. L. Chiarotti, M. Cococcioni, I. Dabo, A. D. Corso, 
\\ \ \ \ \ \ \ \ \ \ \ \ S. Fabris, G. Fratesi, S. de Gironcoli, R. Gebauer, U. Gerstmann, C. 
\\ \ \ \ \ \ \ \ \ \ \ \ Gougoussis, A. Kokalj, M. Lazzeri, L. Martin-Samos, N. Marzari, F. 
\\ \ \ \ \ \ \ \ \ \ \ \ Mauri, R. Mazzarello, S. Paolini, S. Scandolo, G. Sclauzero, A. P. 
\\ \ \ \ \ \ \ \ \ \ \ \ Seitsonen, A. Smogunov, P. Umari, R. M. Wentzcovitch, J. Phys.
\\ \ \ \ \ \ \ \ \ \ \ \ Condens. Matter \textbf{21}, 395502 (2009).

\footnotesize 11) Phonon calculations with the $\bf{q}$-mesh of 6$\times$6$\times$6 so as to check the 
\\ \ \ \ \ \ \ \ \ \ \ \ $\bf{q}$-mesh dependence have not been completed due to CPU time consu-
\\ \ \ \ \ \ \ \ \ \ \ \ ming but have yielded some preliminary results of the electron-phon-
\\ \ \ \ \ \ \ \ \ \ \ \ on coupling constants for several $\bf{q}$-points wihch have been confirmed
\\ \ \ \ \ \ \ \ \ \ \ \ to almost coincide with those for the same $\bf{q}$-points obtained by the 
\\ \ \ \ \ \ \ \ \ \ \ \ present phonon calculations with the $\bf{q}$-mesh of 4$\times$4$\times$4.

\footnotesize 12) H. J. Monkhorst, J. D. Pack, Phys. Rev. B \textbf{13}, 5188 (1976).

\footnotesize 13) B. Fauqu\'{e}, C. Collignon, H. Yoon, Ravi, X. Lin, I. I. Mazin, H. Y. 
\\ \ \ \ \ \ \ \ \ \ \ \  Hwang, K. Behnia Phys. Rev. Res. \textbf{5}, 033080 (2023).

\footnotesize 14) A. Jain, S. P. Ong, G. Hautier, W. Chen, W. D. Richards, S. Dacek, S. 
\\ \ \ \ \ \ \ \ \ \ \ \ Cholia, D. Gunter, D. Skinner, G. Ceder, and K. A. Persson, APL Ma-
\\ \ \ \ \ \ \ \ \ \ \ \ ter. \textbf{1}, 011002 (2013).

\footnotesize 15) W. I. McMillan, Phys. Rev. \textbf{167}, 331 (1968).

\footnotesize 16) P. B. Allen and R. C. Dynes, Phys. Rev. B \textbf{12}, 905 (1975). 
 
\footnotesize 17) C. Collignon, P.Bourges, B. Fauque and K.Behnia, Phys. Rev. X \textbf{10}, 
\\ \ \ \ \ \ \ \ \ \ \ \ 031025 (2020).

\end{document}